\documentclass{article}
\usepackage{spconf,amsmath,graphicx,amsfonts,bm,balance}
\usepackage[square,sort&compress,numbers]{natbib}
\def\x{{\boldsymbol x}}
\def\a{{\boldsymbol a}}
\def\y{{\boldsymbol y}}
\def\Y{{\boldsymbol Y}}
\def\M{{\boldsymbol M}}
\def\A{{\boldsymbol A}}
\def\I{{\boldsymbol I}}

\title{Spectral Method for Multiplexed Phase Retrieval\\ and Application in Optical imaging in Complex Media}
\name{Jonathan Dong$^{1,2}$ \qquad Florent Krzakala$^2$ \qquad Sylvain Gigan$^1$}
\address{
\normalsize $^1$ Laboratoire Kastler Brossel, ENS-Universit\'e PSL, CNRS, Sorbonne Universit\'e, \\ \normalsize Coll\`ege de France, 24 rue Lhomond, 75005 Paris, France\\
\normalsize $^2$ Laboratoire de Physique Statistique, ENS-Universit\'e PSL, CNRS,\\ \normalsize Sorbonne Universit\'e, 75005, Paris, France
}
\begin{document}
%\ninept
%
\maketitle
\begin{abstract}
We introduce a generalized version of phase retrieval called multiplexed phase retrieval. We want to recover the phase of amplitude-only measurements from linear combinations of them. This corresponds to the case in which multiple incoherent sources are sampled jointly, and one would like to recover their individual contributions. We show that a recent spectral method developed for phase retrieval can be generalized to this setting, and that its performance follows a phase transition behavior. We apply this new technique to light focusing at depth in a complex medium. Experimentally, although we only have access to the sum of the intensities on multiple targets,  we are able to separately focus on each ones, thus opening potential applications in deep fluorescence imaging and light delivery. 
\end{abstract}
\section{Introduction}
\label{sec:intro}

To recover a complex-valued object from amplitude measurements only is a computational problem known as phase retrieval. The first iterative algorithms date back to the works of Gerchberg and Saxton \cite{gerchberg1972rw} followed by Fienup \cite{fienup1982phase}, with very successful applications in optics such as the measurement and correction of aberrations in the space telescope Hubble in 1990 \cite{fienup1993hubble}. Since then, phase retrieval algorithms have been applied in a number of domains including microscopy \cite{miao2008extending}, astronomy \cite{fienup1987phase}, acoustics \cite{balan2006signal}, and quantum mechanics \cite{corbett2006pauli}.

Recently, new concepts have been introduced to solve the phase retrieval problem. Semi-definite programming algorithms \cite{candes2013phaselift, waldspurger2015phase} are guaranteed to converge but computationally and memory intensive, while non-linear optimization methods \cite{netrapalli2013phase, candes2015phase, chen2015solving} are very efficient for real-world applications. In all these algorithms, there is an initialization step, which consists in finding a good initial guess for subsequent iterative algorithms. For this purpose, a spectral method has been proposed \cite{netrapalli2013phase, candes2015phase}, setting the initial estimate to be the principal eigenvector of a certain matrix (the eigenvector corresponding to the largest eigenvalue). One can even prove that this estimate becomes very close to the optimal solution when the number of measurements is large enough, provided the measurements are random and independent. 

We introduce here the more difficult problem of multiplexed phase retrieval. Instead of measuring the intensity of a signal, we measure the linear combination of the intensities of {\it several} signals. This problem arises for instances when measurements acquired by a physical detector come from several indistinguishable sources. By solving the multiplexed phase retrieval problem, one can 
%potentially 
unmix the signal from each source and ideally retrieve information about each of the sources. 

In this paper, we show that this multiplexed phase retrieval problem is well suited for spectral methods. When the number of measurement is large enough, the leading eigenvector retrieves the signal from the brightest source, the second leading eigenvector corresponds to the second brightest source, and so on. We present numerical results that point to a phase transition behavior, similar to the one seen in many estimation problems with spiked matrices \cite{baik2005phase,lesieur2015mmse,perry2016optimality,lesieur2017constrained,lu2017phase,mondelli2017fundamental} in random matrix theory. 

We then experimentally apply the method to complex media optics \cite{rotter2017light}: we show in a proof-of-concept experiment that, by solving a multiplexed phase retrieval problem, we are able to disentangle the signals coming from multiple sources, that could for instance be multiple fluorescent beads buried inside a diffusive material.
% To image deep inside biological tissues requires to invert the effect of the complex light propagation inside the tissue using some feedback loop [REF]. 
%When the feedback signal comes from several sources, they are summed and mixed on the detector. 

The paper is organized as follows: In Section \ref{sec:multiplexed phase retrieval}, we introduce the multiplexed phase retrieval problem. To solve it, we present a spectral algorithm in Section \ref{sec:spectral method} and numerical results in Section \ref{sec:results}. Finally, we present an application for imaging in complex media in Section \ref{sec:optics}.

\section{Multiplexed phase retrieval}
\label{sec:multiplexed phase retrieval}

\begin{figure*}[t]
    \begin{center}
        \includegraphics[width=0.75\linewidth]{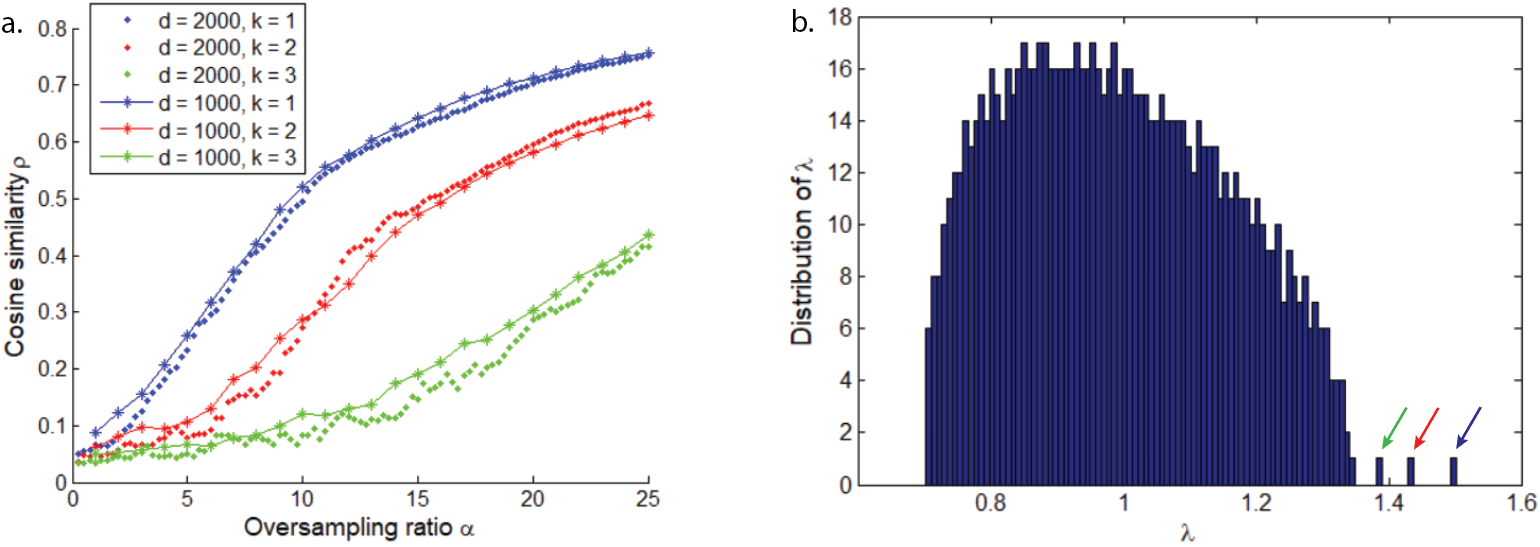}
    \end{center}
    \caption{Phase transitions in spectral methods for multiplexed phase retrieval. \textbf{a.} Cosine similarity $\rho$ as a function of the oversampling ratio $\alpha$ (higher is better). Three sources are present with respective weights $\frac{5}{12}$ (in blue), $\frac{4}{12}$ (in red), $\frac{3}{12}$ (in green). Dots correspond to $d = 2000$ and lines to $d = 1000$, each point is a mean of 10 realizations. \textbf{b.} Histogram of eigenvalues of the Weighted Covariance Matrix $\Y$ for $\alpha = 50$.}
    \label{fig:random gaussian}
\end{figure*}

We will first introduce the phase retrieval problem as formulated by \cite{candes2015phase}, followed by the multiplexed version. 
We consider the following set of quadratic equations:
\begin{equation}
    \label{eq:phase retrieval}
    y_i = |\a_i^* \x|^2, \quad i \in \{1, ..., n\}
\end{equation}
where $\x \in \mathbb{C}^d$ is an unknown vector, $\a_i \in \mathbb{C}^d$ are known sampling vectors, and $\y_i \in \mathbb{R}$ are measured intensities. 
Without the modulus square operation, this set of equations would boil down to a linear regression. Here, we need to retrieve the phase of $\a_i^* \x$ in order to find $\x$. Let us define the sampling matrix $\A = [\a_1^*, ..., \a_n^*] \in \mathbb{C}^{n \times d}$.

% We are going to concatenate the sampling vectors into a sampling matrix $\A = [\a_1^*, ..., \a_n^*] \in \mathbb{C}^{n \times d}$ and the measurements into a vector $\y = [y_1 ... y_n] \in \mathbb{C}^{n \times 1}$. The previous set of equations then writes:
% \begin{equation}
%     \label{eq:matricial phase retrieval}
%     \y = |\A \x|^2
% \end{equation}
% where the modulus operation is taken as a component-wise operation.

% Phase retrieval appears in many different applications, which differs in the sampling matrix $\A$. In many optical applications [REF], $\A$ is a Fourier matrix, on the other hand, w
We will consider the case where $\A$ is a random matrix: every component is an independent random variable, following for example a complex gaussian law: $a_{i,j} \sim_{i.i.d.} \mathcal{CN}\left(0, \frac{1}{d}\right)$. Thus, every sampling vector $\a_i$ is an independent realization of a random vector $\a$. Reconstruction performance for this random model has been the subject of extensive theoretical studies \cite{lu2017phase, mondelli2017fundamental}. One important parameter is the oversampling ratio $\alpha = \frac{n}{d}$, which quantifies the difficulty of the problem. A higher oversampling ratio corresponds to a simpler phase retrieval problem, as we acquire more information about $\x$.

In multiplexed phase retrieval, we want to retrieve several normalized orthogonal vectors $\x_1$, ..., $\x_K$ from a linear combination of intensity measurements:
\begin{equation}
    y_i = \sum_{k=1}^K \lambda_k |\a_i \x_k|^2, \quad i \in \{1, ..., n\}
    \label{eq:multiplexed phase retrieval}
\end{equation}
For $k = 1, ..., K$, each signal $\x_k$ corresponds to a source with strength $\lambda_k$. For normalization, we will assume that $\sum_k \lambda_k = 1$. In optics, a sum of intensities is linked with incoherence, it occurs for instance when several incoherent sources are present. Fundamentally, multiplexed phase retrieval tackles the problem when these sources are physically indistinguishable as they are measured using the same sampling matrix. This situation arises in complex media optical imaging as explained in section \ref{sec:optics}. 
% As a side note, multiplexed Fourier Ptychographic Microscopy \cite{tian2014multiplexed} considers the other case where the object is fixed and intensities from multiple sampling vectors are summed together. 

The multiplexed phase retrieval problem is not always solvable. We can write the measured intensities as:
\begin{equation}
    \label{eq: matricial multiplexed phase retrieval}
    y_i = \a_i^* \M \a_i, \quad i \in \{1, ..., n\} 
\end{equation}
% The link between $\y$ and $\x_k$ is contained in $\M$. 
This matrix $\M = \sum_{k=1}^K \lambda_k \x_k \x_k^*$ defines the measurements $\y$, so the best one can do is to retrieve $\M$, and from there recover the individual signals $\x$. This is only possible if $\M$ has no degenerate eigenvalue that would mix the eigenvectors, i.e. if all the $\lambda_k$ are pairwise distinct. Note also that orthogonality between the $\x_k$ is important as the eigenvectors computed by Support Vector Decomposition form an orthonormal basis. 

\section{Spectral method}
\label{sec:spectral method}

To solve this multiplexed phase retrieval problem, we adapt the spectral method introduced in \cite{lu2017phase, mondelli2017fundamental}. We construct the same weighted covariance matrix and prove that the first few eigenvectors will retrieve each source separately. 

We define the following weighted covariance matrix (WCM):
\begin{equation}
    \Y = \frac{1}{n} \sum_{i=1}^n y_i \a_i \a_i^*
    \label{eq: weighted covariance matrix}
\end{equation}
This WCM is a sum of rank 1 matrices constructed from the sampling vectors, ponderated by the intensity measured for each sampling vector. Without these weights, this matrix would correspond to an empirical estimate of the covariance matrix of the random vector $\a$, and contain no information about $\x_k$ for $k = 1, ..., K$.

Informally, when $\a_i$ is aligned with a certain $\x_k$,  $\y_i$ is higher, so that the WCM is biased towards every $\x_k$. This intuitive argument can be formalized by observing that in the limit $n$ going to infinity, $\Y$ converges towards the expected value of $y \a \a^*$, thanks to the Central Limit Theorem. This asymptotic limit can be computed explicitly by developing $\Y$ as a polynomial of order 4 in the components of $\a$ and computing the expected value of each term separately. When every component of the sampling matrix $\A$ follows a complex gaussian distribution, this expected value can be expressed as:
\begin{equation}
    \overline{\Y} = \M + \I
    \label{eq: expectation gaussian WCM}
\end{equation}
Hence, for a high enough oversampling ratio, the leading eigenvectors of the WCM are going to align with the eigenvectors of $\M$ with eigenvalues $\lambda_k + 1$, while $d-K$ remaining eigenvalues will converge to 1 and form a bulk distribution in Fig. \ref{fig:random gaussian}.

One may wonder how many measurements one needs to perform in order to recover $\M$ using this spectral method. In \cite{candes2015phase}, the authors have observed a phase transition when applying this spectral method in the non-multiplexed phase retrieval problem. When the oversampling ratio $\alpha$ is below a critical value, the leading eigenvectors of $\y$ are random and contain no information about $\M$. Above this critical value, spectral methods retrieve a coarse estimate of $\M$ that converges towards the solution as $\alpha$ grows. This empirical observation has been confirmed for phase retrieval in a theoretical study in \cite{lu2017phase}. In the next section, we will show numerical results supporting a phase transition in the multiplexed case as well.

\section{Results}
\label{sec:results}

In order to evaluate the performance of the previously introduced spectral method in the multiplexed case, we consider the cosine similarity function: $\rho(\x, \y) = \frac{|\x^* \y|}{\|x\|\|y\|}$. It corresponds to the absolute value of the cosine between the two vectors $\x$ and $\y$. As such, this quantity is maximal and equal to 1 when $\x$ and $\y$ are completely aligned, and tends to zero when $\x$ and $\y$ are uncorrelated. 

Figure \ref{fig:random gaussian}a shows the performance of the spectral algorithm as a function of the oversampling parameter $\alpha$. Three vectors $\x_1$, $\x_2$, and $\x_3$ are sampled using a gaussian random matrix $\A$ and their intensities are combined with coefficients $\lambda_1 = \frac{5}{12}$, $\lambda_2 = \frac{4}{12}$, and $\lambda_3 = \frac{3}{12}$. Curves corresponding to $d = 1000$ and $d = 2000$ are superposed, showing that the quantity of interest is indeed the oversampling parameter $\alpha$. We retrieve sources one by one from the brightest to the weakest. We believe that there is a succession of three phase transitions, that are smoothed here due to finite size effects. 

The eigenvalue distribution of the WCM behaves like a spiked random matrix. When $\alpha$ is below the critical value for all the phase transitions, the eigenvalues of $\Y$ lie in a compact support and form a bulk distribution, which is typical for random matrices. During each phase transition, a single eigenvalue comes out and form a spike in the spectrum of $\Y$. When $\alpha$ is large, we see three spikes isolated from the remaining eigenvalues. Hence, looking at the spectrum, we are able to count the number of sources $K$ in the multiplexed phase retrieval problem. 

\section{Imaging in complex media}
\label{sec:optics}

When light propagates in complex media like biological tissues, it encounters many refractive index inhomogeneities and gets scattered multiple times along propagation. This prevents imaging in depth: ballistic light is exponentially attenuated, which limits the penetration depth of conventional microscopy. 

Using techniques of wavefront shaping \cite{vellekoop2007focusing, horstmeyer2015guidestar}, it is now possible to modulate the incoming light precisely in order to form a focus deep inside the scattering medium. These techniques typically use iterative optimization using a feedback signal from a target in order to achieve a focus. It has been in particular shown that by optimizing the incident wavefront of a laser on a scattering medium containing a single buried fluorescent bead, one can focus on it by optimizing the total linear fluorescence signal \cite{van2011optimal}. However, when multiple fluorescent sources are present at depth in a scattering medium, their signals are summed and mixed on the detector, and they cannot be located nor separated. It is possible to harness nonlinearities to form a focus \cite{katz2014noninvasive}, but these methods are not applicable to linear fluorescence. We would like to distinguish them to focus on each of them individually. This task can be formulated as a multiplexed phase retrieval problem. 

\begin{figure}[ht]
    \begin{center}
        \includegraphics[width=\linewidth]{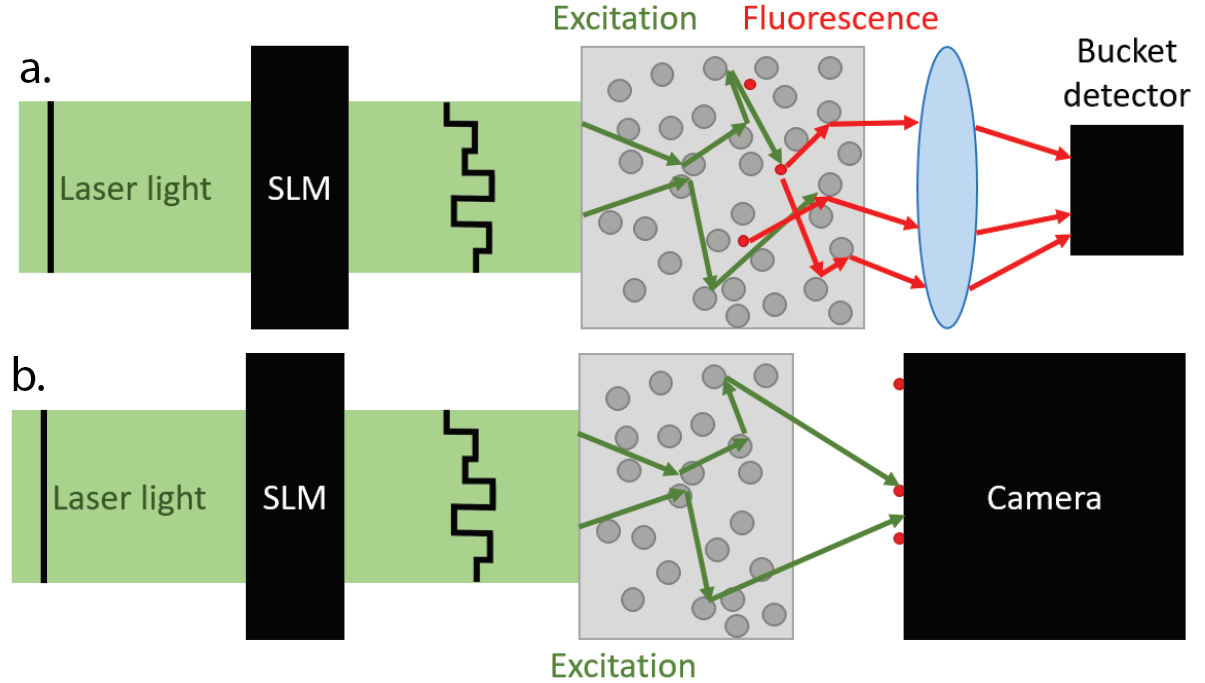}
    \end{center}
    \caption{Focusing light in a scattering media using linear fluorescence signals. \textbf{a.} When multiple fluorescent beads are placed in a complex media, their responses are summed on a bucket detector. \textbf{b.} To study how to focus on each target individually, we sum on a camera the excitation intensity after propagation in a complex media.}
    \label{fig:setup}
\end{figure}

%However in certain applications, feedback signal comes from several sources and wavefront shaping techniques are not able to isolate a single source to obtain a focus. This is the case in linear fluorescence microscopy, where we usually concentrate light on several bright fluorescent targets.

The experimental setup is presented in Figure \ref{fig:setup}b: excitation light from a laser (Coherent Sapphire 532-50 CW) is modulated by a Spatial Light Modulator (SLM, Holoeye Pluto-2 NIR), then propagates through a layer of white paint where it gets scattered multiple times. We place a camera (Basler acA1920 - 40um) after the scattering medium and observe a complex interference pattern called speckle. 

\begin{figure*}[t]
    \begin{center}
        \includegraphics[width=0.7\linewidth]{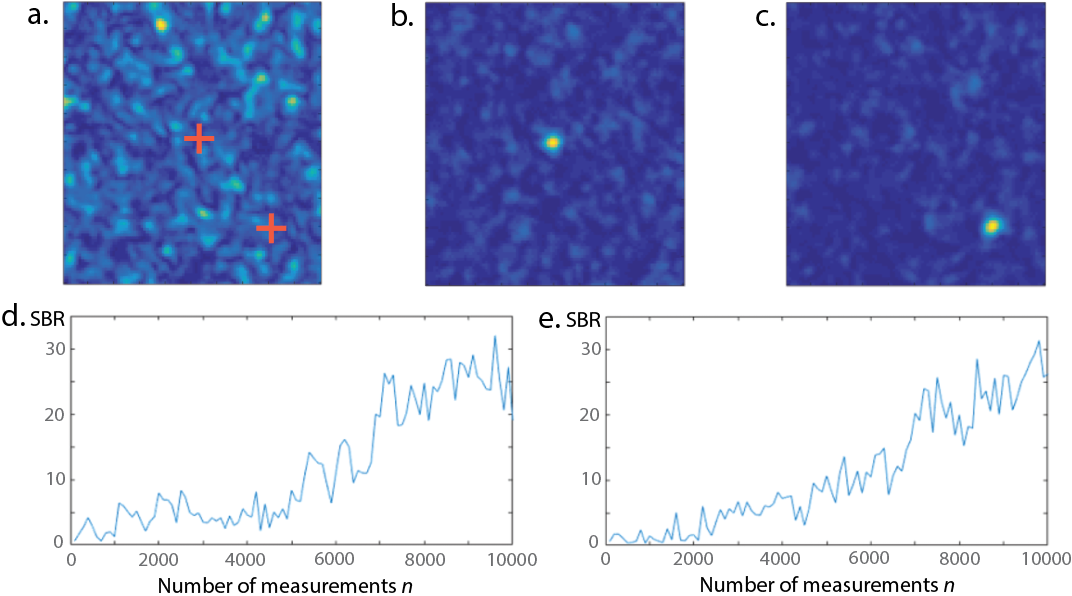}
    \end{center}
    \caption{Experimental application of the spectral method in complex media optics. \textbf{a.} Initial speckle pattern, resulting from light propagation in complex media in the absence of wavefront shaping. We simulate the linear fluorescence signals from targets at the two tagged positions with weights $\lambda_1 = 1$ and $\lambda_2 = 0.7$. \textbf{b, c.} Focus obtained by displaying the phase of the first (b) and second (c) eigenvector of the spectral method. \textbf{d, e.} Signal-to-Background Ratio for the first (d) and second (e) focus as the number of measurements $n$ increases.}
    \label{fig:optics}
\end{figure*}

To show the correspondence, we denote by $\a_i$ the phase pattern displayed on the SLM. We choose two given points on the camera after the scattering medium, and thanks to the Transmission Matrix formalism \cite{popoff2010measuring} that comes from the linearity of electric field propagation, the incident electric field on these two points can be written as $\a_i^* \x_1$ and $\a_i^* \x_2$ respectively, characterized by two vectors $\x_1$ and $\x_2$. Recovering them allows to focus light on each point individually. Fluorescent targets respond proportionally to the excitation intensity, which is the modulus square of the incident electric fields. In the end, we simulate the presence of two fluorescent beads of different brightness, by summing the intensities on these two positions with weights $\lambda_1 = 1$ and $\lambda_2 = 0.7$. We thus collect a signal obeying Equation (\ref{eq:multiplexed phase retrieval}). 

To solve this multiplexed phase retrieval problem, we then send random phase patterns on the SLM and record the corresponding total fluorescence intensity. In this case, the sampling matrix $\A$ is a phase-only random matrix. The dimension of the SLM pattern is $d = 256$ and the maximal number of measurements is $n = 10'000$. Applying the previous spectral method, we obtain estimates of $\x_1$ and $\x_2$ from the WCM, that we can subsequently display on the SLM, and verify whether they correspond to focusing light at the positions where the beads are supposed to be, as shown in Figure \ref{fig:optics}. The oversampling ratio $\alpha = \frac{n}{d}$ is very large to compensate for experimental noise.

% We place a camera after the scattering medium and if $\a_i$ is the phase pattern displayed on the SLM of dimension $d = 256$, we can write the intensity on two given points as $|\a_i^* \x_1|^2$ and $|\a_i^* \x_2|^2$ respectively. This is a consequence of the Transmission Matrix formalism \cite{popoff2010measuring}, that comes from the linearity of the electric field propagation. We then simulate the presence of two fluorescent beads of different brightness, by  selecting two  arbitrary targets positions and summing the intensities on these positions with weights $\lambda_1 = 1$ and $\lambda_2 = 0.7$. We thus collect an intensity obeying Equation (\ref{eq:multiplexed phase retrieval}). 

We see in Figure \ref{fig:optics} experimental results, demonstrating that we are able to focus light on both targets individually with a Signal-to-Background Ratio (SBR) of 25. We believe that a higher SBR is reachable by improving the speed and stability of the experimental setup. 
% In principle, it should scale with the dimension $d = 256$. 
In this first experiment, one experiment takes approximately one hour due to the speed of the SLM and the SBR is probably limited by the stability of the scattering medium. For the same reason, phase transitions appear smoothed because of these experimental considerations.
%probably because the solutions $\x_1$ and $\x_2$ are drifting [NOT CLEAR => intuitively, if the solution is drifting with time, we might not have a phase transition but some averaging effect... Difficult to model and to prove so it's better to avoid it].

\section{Conclusion}
\label{sec:conclusion}

We introduced the multiplexed phase retrieval problem, which arises when multiple signals are incoherently summed. To solve this new task, we applied a spectral method recently designed for phase retrieval initialization. By constructing a Weighted Covariance Matrix and computing its leading eigenvectors, we are able to retrieve the different signals individually. The performance of this algorithm exhibits multiple phase transitions as we increase the number of measurements, recovering each signal one by one. We showed that this framework can be applied to imaging in complex media. Other potential applications where one would like to distinguish between different sources include opto-genetics, photoacoustics and multi-wavelength imaging. 

\section{Acknowledgements}

We thank Leonie Muggenthaler and Romain Couillet for fruitful discussions. SG acknowledges funding from the ERC under the European Union 7th Framework Program Grant Agreements 724473, \& FK from the ANR-PAIL.

%\vfill
%\pagebreak

\bibliographystyle{IEEEbib}
\balance
\bibliography{strings,refs}

\end{document}